\documentstyle[12pt]{article}

   \font\tenmsb=msbm10 scaled\magstep 1
   \font\sevenmsb=msbm7 scaled \magstep 1
   \font\faivemsb=msbm5 scaled \magstep 1
\newfam\msbfam
      \textfont\msbfam=\tenmsb
      \scriptfont\msbfam=\sevenmsb
      \scriptscriptfont\msbfam=\faivemsb
\def\Bbb#1{{\fam\msbfam #1}}
\font\tengothic=eufm10 scaled\magstep 1
\font\sevengothic=eufm7 scaled\magstep 1
\newfam\gothicfam
      \textfont\gothicfam=\tengothic
      \scriptfont\gothicfam=\sevengothic

\newcommand{\Gm}{\Gamma}
\newcommand{\al}{\alpha}
\newcommand{\gm}{\gamma}
\newcommand{\be}{\begin{equation}}
\newcommand{\ee}{\end{equation}}
\newcommand{\dlt}{\delta}
\newcommand{\vp}{\varphi}
\newcommand{\ra}{\rightarrow}
\newcommand{\lbd}{\lambda}
\newcommand{\bt}{\beta}
\newcommand{\om}{\omega}
\newcommand{\prt}{\partial}

\begin{document}

\begin{center}
{\Large{\bf Expansion Exponents for Nonequilibrium Systems} \\ [5mm]
V.I. Yukalov} \\ [3mm]
{\it
Bogolubov Laboratory of Theoretical Physics \\
Joint Institute for Nuclear Research, Dubna 141980, Russia \\
and \\
Research Center for Optics and Photonics \\
Instituto de Fisica de S\~ao Carlos, Universidade de S\~ao Paulo \\
Caixa Postal 369, S\~ao Carlos, S\~ao Paulo 13560-970, Brazil}

\end{center}

\vskip 5cm

{\bf Address for correspondence}:

\vskip 5mm

Prof. V.I. Yukalov

\vskip 2mm

Bogolubov Laboratory of Theoretical Physics 

Joint Institute for Nuclear Research, Dubna 141980, Russia 

\vskip 3mm

{\bf E-mail}: yukalov@thsun1.jinr.ru

\vskip 2mm

{\bf Fax}: 7 (096) 216-5084; $\;\;$ {\bf Phone}: 7 (096) 216-3947  

\newpage

\begin{abstract}

Local expansion exponents for nonequilibrium dynamical systems, 
described by partial differential equations, are introduced. These 
exponents show whether the system phase volume expands, contracts, 
or is conserved in time. The ways of calculating the exponents are 
discussed.  The {\it principle of minimal expansion} provides the 
basis for treating the problem of pattern selection. The exponents 
are also defined for stochastic dynamical systems. The analysis of 
the expansion-exponent behaviour for quasi-isolated systems results 
in the formulation of two other principles: The {\it principle of 
asymptotic expansion} tells that the phase volumes of quasi-isolated 
systems expand at asymptotically large times. The {\it principle of 
time irreversibility} follows from the asymptotic phase expansion, 
since the direction of time arrow can be defined by the asymptotic 
expansion of phase volume.

\end{abstract}

\vskip 3cm


\vskip 1cm


\newpage

\section{Introduction}

Macroscopic properties of nonequilibrium dynamical systems are 
usually characterized by the Gibbs entropy defined for the 
appropriate probability measure [1,2]. The usage of this entropy has 
two shortcomings: First, it is defined only for steady states, and
an effective definition of entropy far from equilibrium is not known
[2]. Second, there appears the problem of singularity resulting from
the following [1,2]. A steady state of a dynamical system corresponds
to the motion on an attractor. The related probability measure is,
typically, singular, because of which the entropy becomes equal to
$-\infty$. To avoid this problem, one considers not the entropy itself 
but the entropy production or entropy production rate, such as the
Kolmogorov-Sinai entropy rate [3]. The latter are finite, but again, 
they are defined only for stationary states.

Employing the Gibbs entropy in nonequilibrium statistical mechanics 
confronts as well another paradox. This entropy, for an isolated system, 
as is well known, does not change in time. While the second law of 
thermodynamics requires that the entropy of an isolated nonequilibrium
system would be an increasing function of time. To overcome this 
contradiction, one introduces modified variants of entropy, among which
the most popular is a coarse-grained enropy [4]. Other variants are the
von Neumann entropy [5] and the average entropy [6], which change with
time for an isolated system not in equilibrium. Lebowitz [7] argues that
the relevant entropy for understanding the time evolution of macroscopic
systems is not the Gibbs entropy $S_G\equiv -{\rm Tr}\rho\ln\rho$ but the
Boltzmann entropy $S_B\equiv\ln|\Gm|$, where $\Gm$ is the system phase 
volume. Contrary to the Gibbs entropy, which is constant for an isolated
macroscopic system, the Boltzmann entropy should typically increase
with time. A weak point in resorting to the Boltzmann entropy $S_B$ is
that it is not always clear how to determine the phase volume for a 
strongly nonequilibrium system.

Instead of trying to construct a special form of entropy that would be
valid for nonequilibrium dynamical systems, it is possible to employ
another characteristic. The aim of the present paper is to demonstrate 
that the {\it local expansion exponent} can be such a convenient 
quantity characterizing the temporal evolution of dynamical systems. 
In Section 2, the local expansion exponent is introduced for an 
infinite-dimensional dynamical system and its properties are described. 
Section 3 shows the usefulness of the expansion exponents for treating
the problem of pattern selection. The generalization of the expansion
exponents to stochastic dynamical systems is given in Section 4.
Several examples of Section 5 illustrate how the exponents can be
calculated. The notion of quasi-isolated systems is formulated in
Section 6 and some particular cases are considered. Section 7 contains
conclusions and discussion.

\section{Local expansion exponents}

Let $x\in\Bbb{D}\subset\Bbb{R}^d$ be a $d$-dimensional set of 
continuous variables given on a domain $\Bbb{D}$ and let 
$t\in\Bbb{R}_+\equiv[0,\infty)$ denote time. Suppose we are interested 
in the evolution of a family of complex-valued functions $y_i(x,t)$, 
with $i\in\Bbb{N}_+\equiv\{ 1,2,\ldots\}$, describing a nonequilibrium
physical process. It is convenient to employ the matrix notation [8],
where the pair $\{ i,x\}$ is treated on equal footing as a point in the
{\it label space} $\Bbb{N}_+\times\Bbb{D}$. Then the {\it dynamic state}
$y(t)=[y_i(x,t)]$ is presented as a vector-column with respect to the
multi-index $\{i,x\}$. The components $y_i(x,t)$ may correspond to some 
physical quantities. In particular, these could be statistical or 
quantum averages of operators for local observables. This could also be
a wave function or a density matrix.

Dynamic state $y(t)\in{\cal F}$ pertains to a {\it phase space} 
${\cal F}$. Let the velocity field $v(y,t)=[v_i(x,y,t)]$ be given as
a column with respect to the label pair $\{ i,x\}$ and defined in a 
space tangent to ${\cal F}$. The set of evolution equations can be 
compactly written as
\be
\label{1}
\frac{d}{dt}\; y(t) = v(y,t) \; .
\ee
For the dynamical components $y_i(x,t)\in{\cal F}_i(x)$ pertaining to
the phase subspaces ${\cal F}_i(x)$, the structure of the total phase
space ${\cal F}$ is that of the tensor product
$$
{\cal F} =\otimes_i \otimes_x {\cal F}_i(x) \; .
$$
The continuous product of spaces ${\cal F}_i(x)$, with respect to the
continuous variable $x$, is a direct generalization [9] of the discrete
product over the discrete index $i=1,2,\ldots$. The related continuous 
product of functions $f(x)$ is naturally defined [9] as
$$
\prod_x f(x) \equiv \exp \int \ln f(x)\; dx \; ,
$$
with the integration over $\Bbb{D}$, where any branch of the logarithm 
may be taken. The {\it elementary phase volume} at time $t$ is 
interpreted [10,11] as
\be
\label{2}
\dlt\Gm(t) \equiv \prod_i \prod_x \dlt y_i(x,t) \; .
\ee

Introduce the {\it local expansion exponent}
\be
\label{3}
X(t) \equiv \ln\left | \frac{\dlt\Gm(t)}{\dlt\Gm(0)}\right | \; ,
\ee
which shows whether the phase volume, at time $t$, expands, contracts,
or is conserved. Rewriting the definition (3) in the form
\be
\label{4}
|\dlt\Gm(t)| = |\dlt\Gm(0)|\; e^{X(t)} \; ,
\ee
one sees that positive $X(t)$ implies expansion, negative $X(t)$ means
contraction, while zero $X(t)$ tells that the phase volume at time $t$ 
does not change. Note that $X(t)$ could equally be called the local 
contraction exponent. However, to term it the expansion exponent looks
more logical as far as $X(t)>0$ implies expansion.

One may notice that the quantity (3) resembles the entropy production
$S(t)-S(0)$, with the entropy $S(t)=\ln|\dlt\Gm(t)|$ defining \`a la
Boltzmann. But, clearly, this is just a resemblance since the expansion 
exponent (3) has sense for any dynamical system and does not require to
be ascribed to a statistical system. For the latter, one usually
defines the entropy production being based on the Gibbs entropy, which
allows for the description of only steady states [1,2]. But for strongly
nonequilibrium systems the Boltzmann form is more appropriate [7].

The expansion exponent (3) is directly related to the multiplier matrix
$\hat M(t)=[M_{ij}(x,x',t)]$ with the elements
\be
\label{5}
M_{ij}(x,x',t) \equiv \frac{\dlt y_i(x,t)}{\dlt y_j(x',0)} 
\ee
and with the initial condition
\be
\label{6}
M_{ij}(x,x',0) = \dlt_{ij}\;\dlt(x-x') \; .
\ee
Invoking equations (2) and (5), we have
\be
\label{7}
X(t) = \sum_i \int \ln|M_{ii}(x,x,t)|\; dx \; .
\ee
This expression can be rewritten in a general form not depending on a 
particular representation. For this purpose, let us recall that for
a matrix $\hat A$ the equality
\be
\label{8}
{\rm Tr}\ln\hat A = \ln{\rm det}\hat A 
\ee
is valid. The proof of this equality is based on the fact that the trace
of a matrix does not depend on a matrix representation. Then, employing 
the diagonal representation for $\hat A=[\dlt_{mn} A_n]$, one has
$$
{\rm Tr}\ln\hat A = \sum_n \ln A_n = \ln \prod_n A_n = \ln{\rm det}
\hat A \; ,
$$
which proves equality (8). Defining the matrix
$$
\ln|\hat M(t)| \equiv [ \ln| \; M_{ij}(x,x',t)| ] \; ,
$$
equation (7) may be presented as
\be
\label{9}
X(t) ={\rm Tr}\ln |\hat M(t)| \; .
\ee
And using equality (8), we obtain
\be
\label{10}
X(t) =\ln|{\rm det}\hat M(t)| \; .
\ee
Thus, the expansion exponent (3) is expressed through the multiplier 
matrix (5).

Another expression for the expansion exponent can be found by invoking 
the evolution equation (1). Define the Jacobian matrix 
$\hat J(t)=[J_{ij}(x,x',t)]$ with the elements
\be
\label{11}
J_{ij}(x,x',t) \equiv \frac{\dlt v_i(x,y,t)}{\dlt y_j(x',t)} \; .
\ee
The variation of equation (1) over the initial conditions results in 
the equation
\be
\label{12}
\frac{d}{dt} \; \hat M(t) = \hat J(t)\hat M(t)
\ee
for the multiplier matrix. The initial condition for equation (12),
according to equality (6), is
\be
\label{13}
\hat M(0) =\hat 1 \equiv [\dlt_{ij}\;\dlt(x-x')]\; .
\ee
With the usage of equation (12), we have
$$
\frac{d}{dt}\;{\rm Tr}\ln \hat M(t) = {\rm Tr}\hat M^{-1}(t)\;
\frac{d}{dt}\; \hat M(t) = {\rm Tr}\hat M^{-1}(t) \hat J(t) \hat M(t)=
{\rm Tr}\hat J(t) \; ,
$$
from where it follows that
$$
\frac{d}{dt}\; {\rm Tr}\ln \hat M(t) = {\rm Tr} \hat J(t) \; .
$$
Integrating this equation gives
$$
{\rm Tr}\ln \hat M(t) = \int_0^t {\rm Tr} \hat J(t')\; dt'\; .
$$
And equality (8) yields
$$
{\rm Tr} \ln \hat M(t) = \ln {\rm det} \hat M (t) \; .
$$
Comparing the latter two equations results in the equality
\be
\label{14}
{\rm det} \hat M(t)  =\exp\left\{ \int_0^t {\rm Tr}
\hat J(t')\; dt'\right\} \; .
\ee
Substituting expressions (14) into equation (10) provides us with the 
local expansion exponent
\be
\label{15}
X(t) = {\rm Re} \int_0^t {\rm Tr} \hat J(t')\; dt' \; .
\ee
Equations (10) and (15) are the main forms for calculating the expansion 
exponent.

For practical calculational purposes, it is possible to use different
representations connected with a chosen basis $\{\vp_n(t)\}$ of the
vector-columns $\vp_n(t)=[\vp_{ni}(x,t)]$. In what follows, such a basis 
is assumed to be orthonormal and complete,
$$
\vp_m^+(t)\vp_n(t) = \dlt_{mn} \; , \qquad
\sum_n \vp_n(t) \vp_n^+(t) = \hat 1 \; .
$$
From here, we get a useful equality
\be
\label{16}
\frac{d\vp_m^+(t)}{dt}\; \vp_n(t) + \vp_m^+\;
\frac{d\vp_n(t)}{dt}  = 0 
\ee
for any indices $m$ and $n$. The $n$-representations for the multiplier 
and Jacobian matrices are defined by the elements 
\be
\label{17}
M_{mn}(t) \equiv \vp_m^+(t)\hat M(t)\vp_n(t) \; , \qquad
J_{mn}(t) \equiv \vp_m^+(t) \hat J(t) \vp_n(t) \; .
\ee
These elements are connected with each other through equation (12)
yielding
\be
\label{18}
\frac{d}{dt}\; M_{mn}(t) = \sum_k\left [ J_{mk}(t) M_{kn}(t) +
M_{mk}(t)\vp_k^+(t)\; \frac{d\vp_n(t)}{dt} \; - \vp_m^+(t)\;
\frac{d\vp_k(t)}{dt}\; M_{kn}(t)\right ] \; ,
\ee
where equality (16) has been invoked. The initial condition for equation
(18) is $M_{mn}(0)=\dlt_{mn}$. The $n$-representation can be exploited 
for deriving several important properties.

\vskip 2mm

{\bf Proposition 1}. If the multiplier matrix $\hat M(t)$ possesses the 
eigenvalues $\mu_n(t)$, given by the eigenproblem
\be
\label{19}
\hat M(t) \vp_n(t) = \mu_n(t) \vp_n(t) \; ,
\ee
with the eigenvectors composing a complete orthonormal basis, then the
eigenvalues
\be
\label{20}
\mu_n(t) = \exp\left\{ \int_0^t J_{nn}(t')\; dt'\right\}
\ee
are expressed through the diagonal elements $J_{nn}(t)$ of the Jacobian
matrix.

\vskip 2mm

Proof is given in Appendix A. Note that the Jacobian matrix is, in 
general, nondiagonal.

\vskip 2mm

{\bf Proposition 2}. Suppose a complete orthonormal basis $\{\vp_n(t)\}$
is such that
\be
\label{21}
\vp_m^+(t)\; \frac{d\vp_n(t)}{dt} = 0 \qquad (m\neq n) \; .
\ee
Then $\vp_n(t)$ are the eigenvectors of the multiplier matrix 
$\hat M(t)$ if and only if they are also eigenvectors of the Jacobian 
matrix $\hat J(t)$, with the corresponding eigenvalues $\mu_n(t)$ and
$J_n(t)$ connected by the relation
\be
\label{22}
\mu_n(t) =\exp\left\{ \int_0^t J_n(t')\; dt'\right\} \; .
\ee

\vskip 2mm

Proof is in Appendix B. The difference between equations (20) and (22)
is in the form of the Jacobian matrix. In equation (20), $J_{nn}(t)$ is 
a diagonal element of a generally nondiagonal matrix. While in equation
(22), $J_n(t)$ is an eigenvalue of the Jacobian matrix, which hence is
diagonal. In other words, theorem 2 tells that, if condition (21) is
satisfied, then the multiplier matrix is diagonal if and only if the 
Jacobian matrix is diagonal.

\vskip 2mm

{\bf Proposition 3}. Let the Jacobian matrix have the limit
\be
\label{23}
\lim_{t\ra\infty} \hat J(t) = \hat J
\ee
and let $J_n$ be the eigenvalues of $\hat J$. Then the limit
\be
\label{24}
\lim_{t\ra\infty} \; \frac{d}{dt}\; X(t) = \sum_n \lbd_n
\ee
is the sum of the Lyapunov exponents
\be
\label{25}
\lbd_n ={\rm Re} J_n = \lim_{t\ra\infty} \; 
\frac{1}{t}\; \ln|\mu_n(t)| \; .
\ee

\vskip 2mm

Proof is in Appendix C. The sum of the Lyapunov exponents gives the
entropy production rate for stationary states [1,2]. Hence, the 
left-hand side of equation (24) has the meaning of the entropy
production rate. Recall that summing up only positive Lyapunov
exponents leads to the Kolmogorov-Sinai entropy rate [3]. Contrary
to the entropy production, defined on the basis of the Gibbs entropy 
[1,2], the local expansion exponent (3) does not meet the problem of
singular measure and it has sense not only for stationary states but
for any strongly nonequilibrium states.

\section{Problem of pattern selection}

The local expansion exponent turns out to be a crucial notion for 
solving the problem of pattern selection. This problem arises when the
evolution equations for a physical process possess a multiplicity of
solutions corresponding to different spatio-temporal structures [12].
When all these solutions are stable, it is not clear how one could 
distinguish between them and, respectively, what spatio-temporal 
structures should be treated as more preferable. At the same time,
nature does prefer some structures over the others, since some of them
arise more often for the given experimental protocol. This problem of
pattern selection can be solved as follows.

Let the multiplicity of solutions be parametrized by a label $\bt$ 
pertaining to a label manifold $\Bbb{B}=\{\bt\}$. The submanifodl of
$\Bbb{B}$ related to stable solutions is called [12] {\it stability
balloon}. Each dynamic state $y(\bt,t)$, labelled by $\bt$, corresponds
to a particular spatio-temporal pattern. Because of the parametrization
of dynamic states by the label $\bt$, the dependence on this label also
comes through the multiplier matrix $\hat M(\bt,t)$, Jacobian matrix
$\hat J(\bt,t)$, and the local expansion exponent $X(\bt,t)$. Since 
nature distinguishes among possible patterns differentiating them on
more or less preferable, there should exist a probability distribution
classifying these patterns as more or less preferable. Such a probability
distribution of patterns $p(\bt,t)$ can be obtained by minimizing 
an information functional
$$
I_p(t) = \int I_p(\bt,t)\; d\bt \; , 
\qquad \int p(\bt,t)\; d\bt = 1 \; ,
$$
under the normalization condition on the probability distribution, 
where the integration is over the label manifold $\Bbb{B}$.

The information density $I_p(\bt,t)$ can be found from the following
requirements: The existence of an invariant probability measure
\be
\label{26}
p(\bt,t) |\dlt \Gm(\bt,t)| = p(\bt,0) |\dlt \Gm(\bt,0)|
\ee
and the occurence of an invariant information measure
\be
\label{27}
I_p(\bt,t) |\dlt\Gm(\bt,t)| = I_p(\bt,0) |\dlt\Gm(\bt,0) | \; .
\ee
These equalities tell us that the probability of a pattern in an 
elementary phase volume and the amount of information in such a 
phase volume are temporal invariants. We also require that at the 
initial time $t=0$ the information density be given by the standard 
Shannon form
\be
\label{28}
I_p(\bt,0) = p(\bt,0) \ln p(\bt,0) \; .
\ee
Then, from the invariance of the probability measure (26), we have
$$
p(\bt,0) = p(\bt,t) \left | \frac{\dlt\Gm(\bt,t)}{\dlt\Gm(\bt,0)}
\right | \; .
$$
Substituting this into the Shanon information density (28) and 
using the invariance of the information masure (27), we obtain the 
information density of patterns
\be
\label{29}
I_p(\bt,t) = p(\bt,t) \left [ \ln p(\bt,t) + X(\bt,t) \right ] \; .
\ee
Thus, the information functional to be minimized is the {\it pattern
information}
\be
\label{30}
I_p(t) = \int p(\bt,t)\ln p(\bt,t)\; d\bt +
\int p(\bt,t)X(\bt,t)\; d\bt \; .
\ee
The minimization of the pattern information (30), under the 
normalization condition on $p(\bt,t)$, implies the minimization of the 
conditional information
\be
\label{31}
\tilde I_p(t) = I_p(t) + l(t)\left [ \int p(\bt,t)\; d\bt - 1
\right ] \; ,
\ee
in which
$$
l(t) \equiv \ln Z(t) - 1
$$
is a Lagrange multiplier. Minimizing information (31) results in the
{\it pattern distribution}
\be
\label{32}
p(\bt,t) = \frac{1}{Z(t)} \; \exp\left\{ - X(\bt,t)\right\} \; ,
\ee
with the normalization factor
$$
Z(t) = \int \exp\left\{ - X(\bt,t)\right\} d\bt \; .
$$
The expansion exponent, according to equations (10) and (15), can be 
determined by one of the forms
\be
\label{33}
X(\bt,t) =\ln|{\rm det}\hat M(\bt,t)|\; , \qquad
X(\bt,t) = {\rm Re}\int_0^t {\rm Tr}\hat J(\bt,t')\; dt'\; .
\ee
With the first of these forms, the pattern distribution (32) becomes
\be
\label{34}
p(\bt,t) = \frac{Z^{-1}(t)}{|{\rm det}\hat M(\bt,t)|} \; ,
\ee
where the normalization factor is
$$
Z(t) = \int \frac{d\bt}{|{\rm det}\hat M(\bt,t)|} \; .
$$
The expression for the pattern distribution (32) shows that those 
patterns are more preferable whose expansion exponents are smaller.
The {\it most probable pattern} corresponds to the {\it minimal 
expansion exponent},
\be
\label{35}
\max_\bt p(\bt,t) \Longleftrightarrow \min_\bt X(\bt,t) \; ,
\ee
which constitutes the {\it principle of minimal expansion}.

The probabilistic approach to pattern selection was advanced in Refs. 
[10,11], where the pattern distribution was rather postulated by
invoking analogies with statistical mechanics. Here we have shown how 
this pattern distribution can be derived from the minimization of the
pattern information. The probabilistic approach to pattern selection 
was applied to the problem of {\it turbulent photon filamentation} 
[11,13], with theoretical description being in a very good agreement 
with all experimental observations. This approach makes also the basis
for the {\it method of self-similar prediction} [14--16] applied for
analysing and forecasting possible scenarios in the behaviour of 
nonequilibrium complex systems, such as markets. In the latter case,
the scenario distribution of form (34) was employed.

\section{Stochastic dynamical systems}

The notion of expansion exponents can be generalized to stochastic 
dynamical systems. Suppose the evolution equations include a stochastic 
field $\xi(t)=[\xi_i(x,t)]$. Then the dependence on this field enters the
dynamic state $y(\xi,t)=[y_i(x,\xi,t)]$ as well as the velocity field
$v(y,\xi,t)=[v_i(x,y,\xi,t)]$. All these quantities are again treated 
as columns with respect to the multi-index $\{ i,x\}$ spanning the label
space $\Bbb{N}_+ \times\Bbb{D}$. The set of evolution equations acquires 
the form
\be
\label{36}
\frac{d}{dt}\; y(\xi,t) = v(y,\xi,t) \; ,
\ee
with the initial condition
\be
\label{37}
y(\xi,0) = y(0) \; .
\ee
The observable quantities are obtained after averaging over the 
stochastic fields. Denoting the stochastic averaging through the double
angle brackets $\ll\ldots\gg$, we have
\be
\label{38}
y(t) \equiv \; \ll y(\xi,t) \gg \; .
\ee
In the elementary phase volume (2), we now need to set
\be
\label{39}
\dlt y_i(x,t) = \; \ll \dlt y_i(x,\xi,t)\gg \; .
\ee
Then the same definition (3) for the local expansion exponent holds 
true.

Introducing the stochastic multiplier matrix $\hat M(\xi,t)=
[M_{ij}(x,x',\xi,t)]$, with the elements
\be
\label{40}
M_{ij}(x,x',\xi,t) \equiv \frac{\dlt y_i(x,\xi,t)}{\dlt y_j(x',0)}
\ee
and the initial condition
\be
\label{41}
M_{ij}(x,x',\xi,0) = \dlt_{ij}\;\dlt(x-x') \; ,
\ee
we may define the average multiplier matrix
\be
\label{42}
\hat M(t) \equiv \; \ll \hat M(\xi,t)\gg \; .
\ee
Then expressions (7) and (10) for the expansion exponent retain their 
sense, with the multiplier matrix (42).

The stochastic Jacobian matrix is defined as $\hat J(\xi,t)=
[J_{ij}(x,x',\xi,t)]$, whose elements are
\be
\label{43}
J_{ij}(x,x',\xi,t) \equiv 
\frac{\dlt v_i(x,y,\xi,t)}{\dlt y_j(x',\xi,t)} \; .
\ee
The variation of the evolution equation (36) yields the equation
\be
\label{44}
\frac{d}{dt}\; \hat M(\xi,t) = \hat J(\xi,t)\hat M(\xi,t)
\ee
for the stochastic multiplier matrix (40), the initial condition being
\be
\label{45}
\hat M(\xi,0) = \hat 1 \; .
\ee
The averaging of equation (44) gives
$$
\frac{d}{dt}\; \hat M(t) = \; \ll \hat J(\xi,t)\hat M(\xi,t)\gg \; .
$$
Since the right-hand side here, in general, cannot be factorized, 
there is no presentation for the expansion exponent similar to 
equation (15). Therefore, for stochastic dynamical systems, the 
form (10) of the expansion exponent remains the main, where the 
multiplier matrix is given by equation (42).

With the help of an orthonormal complete basis $\{\vp_n(t)\}$, we may 
pass to the $n$- representation
\be
\label{46}
M_{mn}(\xi,t) \equiv \vp_m^+\hat M(\xi,t)\vp_n(t) \; , \qquad
J_{mn}(\xi,t) \equiv \vp_m^+(t)\hat J(\xi,t) \vp_n(t) \; ,
\ee
which is analogous to definition (17). The elements (46) are connected
by the same equation (18). If the eigenproblem
\be
\label{47}
\hat M(\xi,t)\vp_n(t) = \mu_n(\xi,t)\vp_n(t)
\ee
holds true, then the average multiplier matrix (42) possesses the same 
eigenvectors $\vp_n(t)$, with the eigenvalues
\be
\label{48}
\mu_n(t) = \; \ll \mu_n(\xi,t)\gg \; .
\ee
It is straightforward to reformulate the theorems 1 and 2 for the 
stochastic multiplier matrix $\hat M(\xi,t)$:

\vskip 2mm

{\bf Proposition 1}. If the stochastic multiplier matrix $\hat M(\xi,t)$ 
possesses the eigenvalues $\mu_n(\xi,t)$, given by the eigenproblem 
(47), with the eigenvectors forming a complete orthonormal basis, then 
the eigenvalues
\be
\label{49}
\mu_n(\xi,t) = \exp\left\{ \int_0^t J_{nn}(\xi,t')\; dt'\right\}
\ee
are expressed through the diagonal elements $J_{nn}(\xi,t)$ of the
stochastic Jacobian matrix.

\vskip 2mm

{\bf Proposition 2}. Assume a complete orthonormal basis $\{\vp_n(t)\}$ 
is such that condition (21) is valid. Then the stochastic multiplier 
matrix $\hat M(\xi,t)$ is diagonal in the $n$-representation if and 
only if the stochastic Jacobian matrix is diagonal,
$$
M_{mn}(\xi,t) = \mu_n(\xi,t)\;\dlt_{mn} \; , \qquad
J_{mn}(\xi,t) = J_n(\xi,t)\;\dlt_{mn} \; ,
$$ their eigenvalues being connected by the relation (49).

But theorem 3 is not valid for stochastic dynamical systems, since
$\hat J(\xi,t)$, in general, is not defined for $t\ra\infty$. Thus,
to find the expansion exponent, we have to resort to formula (10). 
When the stochastic multiplier matrix satisfies the eigenproblem (47),
the expansion exponent (10) reduces to the expression
\be
\label{50}
X(t) = \sum_n \ln|\ll \mu_n(\xi,t)\gg | \; .
\ee
This is the principal form of the expansion exponent for stochastic
dynamical systems.

\section{Exponents for stochastic systems}

Now we shall illustrate how the expansion exponents can be calculated 
for several examples of stochastic dynamical systems. We shall consider 
the stochastic field $\xi(t)$ as a real-valued random Gaussian variable,
which is centered at zero,
$$
\ll \xi(t)\gg \; = 0 \; .
$$
For the case of coloured noise, the correlation function
$$
\ll \xi(t)\xi(t') \gg \; = C(t-t')
$$
can be written [17] as
$$
C(t) = C_0\exp\left ( -\; \frac{|t|}{\tau} \right ) \; .
$$
Setting $C_0=(\gm_1/\tau + 2\gm_2^2)$, we may consider two opposite
limits, that of a very short correlation time $\tau$ and of a long 
correlation time. In the short correlation time limit $\tau\ra 0$, 
taking into account that
$$
\frac{1}{2\tau}\; \exp\left ( -\; \frac{|t|}{\tau}\right ) 
\simeq \dlt(t) \qquad (\tau\ra 0) \; ,
$$
we have
$$
\lim_{\tau\ra 0} C(t) = 2\gm_1 \dlt(t) \; .
$$
Such a delta-correlated noise is called white because of the uniformity
of its spectral function. In the opposite limit of a long correlation
time $\tau\ra\infty$, we get
$$
\lim_{\tau\ra\infty} C(t) = 2\gm_2^2 \; .
$$
This kind of noise can be called infrared, since its spectral function 
is centered at zero energy. In the general case, one should deal with
the coloured noise with a finite correlation time. To simplify
calculations, at the same time keeping tracks of two admissible limits 
of short and long correlation times, one can accept the correlation 
function of the form
$$
C(t) = 2\gm_1 \dlt(t) + 2\gm_2^2 \; ,
$$
representing a mixture of white and infrared noises. In what follows, 
we shall consider the Gaussian stochastic fields with the correlation
function
\be
\label{51}
\ll \xi(t) \xi(t') \gg \; = 2\gm_1 \dlt(t-t') + 2\gm_2^2 \; ,
\ee
which includes both limiting cases. Setting here $\gm_1\ra 0$, we get 
the infrared noise, while for $\gm_2\ra 0$, we come to the white noise.

Stochastic differential equations, as is known, can be interpreted 
either in the sense of Ito or in the sense of Stratonovich [17,18].
We prefer to deal with the Stratonovich interpretation, which is
better motivated physically. Another possibility could be to employ 
the stochastic expansion technique [19,20], by presenting stochastic
fields as expansions over smooth functions with random coefficients.
This method allows for the usage of the standard differential and
integration analysis. The final results of the expansion technique
coincide with the corresponding expressions derived by means of the
Stratonovich method.

\subsection{Oscillator with random attenuation}

Let us start with a case of an one-dimensional dynamical system 
described by the evolution equation
\be
\label{52}
\frac{dy}{dt} = \left ( i\om -\Gm +\xi\right ) y \; ,
\ee
in which $y=y(\xi,t)$, $\xi=\xi(t)$, $\om$ and $\Gm$ are positive 
parameters. This equation corresponds to an oscillator with random 
attenuation. The Jacobian related to equation (52) is
\be
\label{53}
J(\xi,t) = i\om -\Gm +\xi(t) \; ,
\ee
because of which the multiplier is
\be
\label{54}
\mu(\xi,t) = \exp\left\{ \left ( i\om - \Gm\right ) t +
\int_0^t \xi(t')\; dt'\right \} \; .
\ee
For the Gaussian random variables, one has the averaging property
$$
\ll \exp\left\{ \al \int_0^t \xi(t')\; dt'\right \} \gg \; =
\exp\left\{ \frac{\al^2}{2} \int_0^t \ll \xi(t')\xi(t'') \gg\;
dt'\; dt'' \right\} \; ,
$$
where $\al$ is a complex number. Then for the local expansion exponent 
(50), we find
\be
\label{55}
X(t) = -\Gm t +\gm_1 t + \gm_2^2 t^2 \; .
\ee
If there is no noise, that is $\gm_1=\gm_2=0$, the motion is 
contracting. But in the presence of noise such that either $\gm_1>\Gm$ 
and $\gm_2=0$ or $\gm_2\neq 0$ for any $\gm_1$, the dynamics is
expanding. As is seen, the existence of noise can drastically change
the system dynamics.

\subsection{Stochastic diffusion equation}

Consider an infinite-dimensional dynamical system presented by the 
diffusion equation with a random diffusion coefficient,
\be
\label{56}
\frac{\prt y}{\prt t} = (D +\xi)\; \frac{\prt^2 y}{\prt x^2} \; ,
\ee
where $y=y(x,\xi,t)$, $\xi=\xi(t)$, and $D>0$. For the spatial variable 
$x$ given on a finite interval, it is always possible, by an appropriate 
scaling, to reduce this interval to the unit one, so that $x\in[0,1]$.
The initial condition is
\be
\label{57}
y(x,\xi,0)=y(x,0) \; .
\ee
And let the boundary conditions be
\be
\label{58}
y(0,\xi,t) = b_0 \; , \qquad y(1,\xi,t) = b_1 \; .
\ee
From here, the boundary conditions for the multiplier matrix are
\be
\label{59}
M(0,x',\xi,t) = M(1,x',\xi,t) = 0 \; .
\ee
Hence, if the multiplier matrix possesses the eigenvectors 
$\vp_n=[\vp_n(x)]$, their components $\vp_n(x)$ should satisfy the 
boundary conditions
\be
\label{60}
\vp_n(0) =\vp_n(1) = 0 \; .
\ee

For the Jacobian matrix, associated with equation (56), we have
\be
\label{61}
J(x,x',\xi) = ( D +\xi)\; \frac{\prt^2}{\prt x^2}\; \dlt(x-x') \; .
\ee
The eigenproblem for the Jacobian matrix reads
$$
\int_0^1 J(x,x',\xi) \vp_n(x')\; dx' = J_n(\xi)\vp_n(x) \; .
$$
The eigenfunctions, satisfying the boundary conditions (60), are
$$
\vp_n(x) =\sqrt{2}\; \sin(\pi n x) \qquad (n=1,2,\ldots) \; .
$$
And the eigenvalues of the Jacobian matrix are
\be
\label{62}
J_n(\xi) = - (D+\xi) \pi^2 n^2 \; .
\ee
The eigenvectors $\vp_n$, not depending on time, satisfy condition 
(21). Then, by theorem 2, the multiplier matrix possesses the same 
eigenvectors, with the eigenvalues (49), which yields
\be
\label{63}
\mu_n(\xi,t) = \exp\left\{ - \pi^2 n^2 Dt - \pi^2 n^2 \int_0^t
\xi(t')\; dt'\right \} \; .
\ee
In this way, the expansion exponent (50) becomes
\be
\label{64}
X(t) = \sum_n \left [ - \pi^2 n^2 Dt + \pi^4 n^4 \left ( \gm_1 t +
\gm_2^2 t^2\right ) \right ] \; .
\ee
Using the sums
$$
\sum_{n=1}^N n^2 = \frac{N}{6}\left ( 2N^2 + 3N + 1\right ) \; , \qquad
\sum_{n=1}^N n^4 = \frac{N}{30}\left ( 6N^4 + 15 N^3 + 10N^2 -1
\right ) \; ,
$$
we obtain
\be
\label{65}
X(t) \simeq -\; \frac{\pi^2}{3}\; N^3 Dt + \frac{\pi^4}{5}\;
N^5 \left ( \gm_1 t + \gm_2^2 t^2 \right ) \; ,
\ee
where $N\ra\infty$. In the absence of noise, when $\gm_1=\gm_2=0$,
the system is contracting. But if either $\gm_1$ or $\gm_2$ are not 
zero, the phase volume expands.

\subsection{Stochastic Schr\"odinger equation}

The Schr\"odinger equation, generated by a Hamiltonian $H$, with added
stochastic background, can be written in the form
\be
\label{66}
\frac{\prt\psi}{\prt t} = (-iH + \xi)\psi \; ,
\ee
which is called normal for dynamical systems. Here $\psi=
\psi({\bf r},\xi,t)$, $\xi=\xi(t)$, $H=H({\bf r})$, and $\hbar=1$.

For the Jacobian matrix, we have
\be
\label{67}
J({\bf r},{\bf r}',\xi) = (-iH+\xi)\;\dlt({\bf r}-{\bf r}') \; .
\ee
Let the stationary Schr\"odinger equation be
$$
H\psi_n = E_n\psi_n \qquad \left ( \sum_n 1 \equiv N\right ) \; .
$$
The Jacobian matrix $\hat J(\xi)$, with the elements (67), possesses
the eigenvectors $\psi=[\psi_n(x)]$ and the eigenvalues
\be
\label{68}
J_n(\xi) = - iE_n + \xi \; .
\ee
Since the eigenvectors $\psi$ do not depend on time, condition 
(21) is valid. Then, by theorem 2, we find the eigenvalues
\be
\label{69}
\mu_n(\xi,t) = \exp\left\{ - i E_n t + \int_0^t \xi(t')\; dt'
\right \} 
\ee
of the multiplier matrix. And for the expansion exponent (50), we
find
\be
\label{70}
X(t) =\left ( \gm_1 t + \gm_2^2 t^2 \right ) N \; .
\ee
When the noise is absent, the phase volume is conserved. But as soon 
as either $\gm_1$ or $\gm_2$ are not zero, the dynamical system expands.

\section{Exponents for quasi-isolated systems}

All real physical systems are never completely isolated, but are
subject to weak uncontrollable random perturbations from surrounding.
This fact has been repeatedly emphasized and discussed in literature 
[6,9,22--25]. Even more, as has been stressed [26,27], the notion of 
an isolated system as such is logically self-contradictory. This is
because to practically realize the isolation one has to use technical
devices acting on the system, and to ensure that the latter is kept
isolated, one must apply measuring instruments perturbing the system. 
The preparation and registration processes may essentially influence
the system evolution [5,28]. The impossibility of isolating macroscopic
systems from their environments is often considered as the cause of the
encrease of entropy required by the second law of thermodynamics and
the related irreversibility of time arrow [28].

To analyze the influence of weak external noise on the system dynamics,
it is useful to introduce the notion of quasi-isolated systems [8]. A
physical system is called quasi-isolated if it is subject to the action
of very small stochastic perturbations modelling the random influence
of surrounding. In order to explicitly show that stochastic perturbations 
are weak, they can be included in the evolution equations with a small 
factor $\al\ll 1$, so that such equations take the form
\be
\label{71}
\frac{d}{dt}\; y(\al\xi,t) = v(y,\al\xi,t) \; .
\ee
As a result, the multiplier matrix $\hat M(\al\xi,t)$ and the Jacobian 
matrix $\hat J(\al\xi,t)$ also contain this small parameter. And the 
local expansion exponent becomes
\be
\label{72}
X(\al,t) = \ln|{\rm det}\; \ll\hat M(\al\xi,t)\gg | \; .
\ee

An isolated system could be treated as the asymptotic limit of the
related quasi-isolated system, as $\al\ra 0$. However, there is a
very gentle point in taking such an asymptotic limit. A system, being
considered for long time, corresponds to the temporal limit $t\ra\infty$.
The limits $\al\ra 0$ and $t\ra\infty$ may turn to be noncommuting!
If these limits for the expansion exponent (72) do not commute, so that
\be
\label{73}
[\lim_{t\ra\infty}, \; \lim_{\al\ra 0}] X(\al,t) \neq 0 \; ,
\ee
then the notion of an isolated system has not much sense, since 
infinitesimally small random perturbations principally change the
system behaviour. Condition (73), when it holds true, can be accepted 
as a mathematical formulation for the principle of {\it nonexistence
of isolated systems}.

For illustration, we may employ the examples of Section 5. Thus, for
the stochastic oscillator of subsection 5.1 we have
\be
\label{74}
X(\al,t) = -\Gm t +\al^2 \left ( \gm_1 t + \gm_2^2 t^2 \right ) \; .
\ee
This gives us the limits
\be
\label{75}
\lim_{t\ra\infty}\lim_{\al\ra 0} X(\al,t) = -\infty \; , \qquad
\lim_{\al\ra 0}\lim_{t\ra\infty} X(\al,t) = +\infty \; ,
\ee
where it is assumed that $\gm_2\neq 0$. For the stochastic diffusion
equation of subsection 5.2, we get
\be
\label{76}
X(\al,t) \simeq -\; \frac{\pi^2}{3}\; N^3 Dt + \al^2 \; 
\frac{\pi^4}{5}\; N^5 \left ( \gm_1 t + \gm_2^2 t^2\right ) \; ,
\ee
from where it follows that
\be
\label{77}
\lim_{t\ra\infty}\lim_{\al\ra 0} X(\al,t) = -\infty \; , \qquad
\lim_{\al\ra 0}\lim_{t\ra\infty} X(\al,t) = +\infty \; .
\ee
And for the stochastic Schr\"odinger equation of subsection 5.3, 
we find
\be
\label{78}
X(\al,t) =\al^2 \left ( \gm_1 t + \gm_2^2 t^2\right ) N \; ,
\ee
which yields the limits
\be
\label{79}
\lim_{t\ra\infty}\lim_{\al\ra 0} X(\al,t) = 0 \; , \qquad
\lim_{\al\ra 0}\lim_{t\ra\infty} X(\al,t) = +\infty \; ,
\ee
As we see, irrespectively of whether the system without random 
perturbation has been phase conserving or contracting, it turns, 
in the long run, to an expanding system as soon as it is influenced 
by a noise, no matter how weak the noise is.

In this way, at finite times the local expansion exponent can be 
negative or zero; for some systems it may, probably, fluctuate,
similarly to entropy fluctuations in macroscopic kinetics [29].

But there always exist such small random perturbations that at
sufficiently long times the expansion exponent becomes positive.
This suggests to formulate the principle of {\it asymptotic phase
expansion},
\be
\label{80}
X(t)> 0 \qquad (t\ra\infty) \; ,
\ee
telling that the expansion exponents of quasi-isolated systems become
positive at asymptotically large times. Thus, any real physical system
can be treated as approximately isolated only for a finite time 
interval. But sooner or later, the influence of its uncontrollable 
stochastic environment will prevail and the system phase volume will
start expanding.

The {\it principle of asymptotic expansion} (80) is explicitly related 
to the direction of time. Therefore, the {\it irreversibility of time 
arrow} can be treated as a consequence of this principle. One often
connects the irreversibility of time with the increase of entropy
postulated for isolated systems by the second law of thermodynamics. 
The increase of entropy could be due to the internal chaotic nature 
of the microscopic dynamics [7]. However, as is recently reviewed 
by Zaslavsky [30], chaotic dynamics in real systems does not provide
finite relaxation time to equilibrium or fast decay of fluctuations, 
and chaotic systems are not completely random in the sense originally
postulated for statistical systems. Chaotic systems possess a property
different from the regular understanding of randomness, a property
called {\it persistence of nonequilibrium} [30]. Thus, chaotic dynamics 
of isolated systems cannot provide an explanation for the increase 
of entropy. Moreover, there is no effective definition of entropy 
for systems far from equilibrium [2]. Contrary to entropy, the 
local expansion exponent is defined not only for statistical systems 
but for arbitrary dynamical systems, strongly as well as weakly 
nonequilibrium. An important fact is that completely isolated systems 
do not exist in nature. Any real system can be only quasi-isolated, 
being subject to uncontrollable random influence of its environment.
Although the internal chaotic dynamics, if any, may play the role, 
but the main cause for the property of asymptotic expansion (80) 
is the influence of stochastic surrounding. It is this asymptotic 
phase expansion that indicates the direction of time and makes the 
time arrow irreversible. The definition of the expansion exponent 
does not invoke any thermodynamic or statistical notions, but is 
valid for arbitrary dynamical systems. This is why the irreversibility 
of time arrow is not a privilege of only statistical and thermodynamic
systems, but it is a common property of all {\it evolution processes}.

\section{Conclusion}

The notion of local expansion exponents is introduced being valid 
for arbitrary dynamical systems, including stochastic dynamical 
systems. This notion plays a fundamental role in the problem of 
pattern selection. The probabilistic approach to pattern selection 
yields the principle of {\it minimal expansion exponent}, according
to which the most probable pattern at a given time corresponds to
the minimal local expansion exponent.

Considering the limits of large times and of weak stochastic sources,
as applied to the expansion exponent, makes it possible to give a
mathematical formulation for the principle of {\it nonexistence of
isolated systems}, when these two limits do not commute with each 
other.

Several examples of quasi-isolated systems suggest, as a plausible 
generalization, the principle of {\it asymptotic phase expansion},
stating that the expansion exponents of quasi-isolated systems become
positive at sufficiently long times.

The positive definiteness of the expansion exponent at asymptotically
large times implies the expansion of the system phase volume. The
relation between the asymptotic phase expansion and time defines the
direction of time and provides the foundation for the principle of
{\it irreversibility of time arrow}, valid for all evolution processes.

\vskip 5mm

{\bf Acknowledgement}

\vskip 2mm

I appreciate financial support from the S\~ao Paulo State Research
Foundation, Brazil and from the Bogolubov-Infeld Program, Poland.

\newpage

{\large{\bf Appendix A}}: Proof of Proposition 1

\vskip 3mm

When the multiplier matrix satisfies the eigenproblem (19), its 
elements, defined in equation (17), are diagonal,
$$
M_{mn}(t) =\dlt_{mn}\;\mu_n(t) \; .
$$
Substituting this into equation (18) gives
$$
\dlt_{mn}\; \frac{d}{dt}\; \mu_n(t) = J_{mn}(t)\mu_n(t) +
\left [ \mu_m(t) - \mu_n(t)\right ] \vp_m^+(t)\;
\frac{d\vp_n(t)}{dt} \; ,
$$
where the property (16) has been used. For $m=n$, we get the equation
$$
\frac{d}{dt}\; \mu_n(t) =  J_{nn}(t)\mu_n(t) \; ,
$$
with the initial condition
$$
\mu_n(0)=1 \; ,
$$
following from condition (13). The solution of the latter equation 
results in the eigenvalues (20) containing the diagonal elements of
the Jacobian matrix. In general, the latter matrix is not diagonal, 
with the nondiagonal elements being
$$
J_{mn}(t) =\left [ 1 -\; \frac{\mu_m(t)}{\mu_n(t)}\right ]
\vp_m^+(t) \; \frac{d\vp_n(t)}{dt} \; ,
$$
where $m\neq n$.

\vskip 5mm

{\large{\bf Appendix B}}: Proof of Proposition 2

\vskip 3mm

If condition (21) is valid, then equation (18) becomes
$$
\frac{d}{dt}\; M_{mn}(t) = \sum_k J_{mk}(t) M_{kn}(t) + 
M_{mn}(t) \left [ \vp_n^+(t)\;
\frac{d\vp_n(t)}{dt} \; - \vp_m^+(t)\; \frac{d\vp_m(t)}{dt}
\right ] \; .
$$

Suppose that $\vp_n(t)$ are the eigenvectors of the multiplier matrix
$\hat M(t)$, so that the eigenproblem (19) holds. Then the above 
equation yields
$$
\dlt_{mn}(t) \; \frac{d}{dt}\; \mu_n(t) = J_{mn}(t)\mu_n(t)\; ,
$$
from where it follows that the Jacobian matrix is diagonal
$$
J_{mn}(t) =\dlt_{mn}J_n(t) \; ,
$$
that is, $\vp_n(t)$ are also the eigenvectors of $\hat J(t)$. The
eigenvalues $\mu_n(t)$ and $J_n(t)$ are related by expression (22).

Let now $\vp_n(t)$ be the eigenvectors of the Jacobian matrix 
$\hat J(t)$. Then equation (18) can be solved as
$$
M_{mn}(t) = M_{mn}(0)\exp\left\{ \int_0^t \left [
J_m(t') + \vp_n^+(t')\; \frac{d\vp_n(t')}{dt'}\; -
\vp_m^+(t')\; \frac{d\vp_m(t')}{dt'} \right ]\; dt'\right\} \; .
$$
From here, with the initial condition $M_{mn}(0)=\dlt_{mn}$, we have
$$
M_{mn}(t)=\dlt_{mn}\exp\left \{ \int_0^t J_n(t')\; dt' \right\} \; ,
$$
which shows that the multiplier matrix is diagonal. Hence, $\vp_n(t)$ 
are also the eigenvectors of $\hat M(t)$, with the eigenvalues (22).

\vskip 5mm

{\large{\bf Appendix C}}: Proof of Proposition 3

\vskip 3mm

Differentiating the local expansion exponent (15) gives
$$
\frac{d}{dt}\; X(t) = {\rm Re}\; {\rm Tr} \hat J(t)
$$
for all $t\geq 0$. Due to the existence of the limit (23),
$$
\lim_{t\ra\infty}\; \frac{d}{dt}\; X(t) = {\rm Re}\; {\rm Tr}
\hat J \; .
$$
With $J_n$ being the eigenvalues of $\hat J$, one has
$$
{\rm Re}\; {\rm Tr} \hat J = \sum_n{\rm Re}\; J_n = \sum_n \lbd_n \; ,
$$
where $\lbd_n\equiv{\rm Re}J_n$ are the Lyapunov exponents. The
eigenvectors of $\hat J$ do not depend on time, hence they satisfy
condition (21). Then, by theorem 2, the limit of $\hat M(t)$ at
$t\ra\infty$ possesses the same eigenvectors as $\hat J$. At
asymptotically large $t$, because of representation (10) one has
$$
X(t) \simeq \sum_n \ln |\mu_n(t)| \qquad (t\ra\infty) \; .
$$
And relation (22) gives
$$
\ln|\mu_n(t)|\simeq {\rm Re}\; \int_0^t J_n(t')\; dt' \qquad
(t\ra\infty) \; .
$$
Therefore
$$
\lim_{t\ra\infty} \; \frac{1}{t}\; \ln|\mu_n(t)| = {\rm Re}\;
J_n =\lbd_n \; .
$$
Thus, we prove equations (24) and (25).


\newpage

\begin{thebibliography}{99}
\bibitem{1}
D. Ruelle, J. Stat. Phys. 85 (1996) 1.

\bibitem{2}
D. Ruelle, J. Stat. Phys. 95 (1999) 393 .

\bibitem{3}
G.M. Zaslavsky, Chaos in Dynamical Systems, Harwood, New York, 1984.

\bibitem{4}
L. Rondoni, E.G.D. Cohen, Nonlinearity 13 (2000) 1905.

\bibitem{5}
J. von Neumann, Mathematical Foundations of Quantum Mechanics,
Princeton University, Princeton, 1955.

\bibitem{6}
V.I. Yukalov, Statistical Green's Functions, Queen's University,
Kingston, 1998.

\bibitem{7}
J.L. Lebowitz, Physica A 263(1999) 516.

\bibitem{8}
V.I. Yukalov, Physica A 234 (1997) 725.

\bibitem{9}
V.I. Yukalov, Phys. Rep. 208 (1991) 395.

\bibitem{10}
V.I. Yukalov, Phys. Lett. A 284 (2001) 91.

\bibitem{11}
V.I. Yukalov, Physica A 291 (2001) 255.

\bibitem{12}
M.C. Cross, P.C. Hohenberg, Rev. Mod. Phys. 65 (1993) 851.

\bibitem{13}
V.I. Yukalov, Phys. Lett. A 278 (2000) 30.

\bibitem{14}
V.I. Yukalov, S. Gluzman, Int. J. Mod. Phys. B 13 (1999) 1463.

\bibitem{15}
V.I. Yukalov, Mod. Phys. Lett. B 14 (2000) 791.

\bibitem{16}
V.I. Yukalov, Eur. Phys. J. B 20 (2001) 609.

\bibitem{17}
C.W. Gardiner, Handbook of Stochastic Methods, Springer, Berlin,
1997.

\bibitem{18}
R.L. Stratonovich, Topics in the Theory of Random Noise, Gordon 
and Breach, New York, 1963.

\bibitem{19}
R.L. Stratonovich, Nonlinear Nonequilibrium Thermodynamics, 
Springer, Berlin, 1992.

\bibitem{20}
V.I. Yukalov, Phys. Rev. A 56 (1997) 5004.

\bibitem{21}
V.I. Yukalov, Laser Phys. 7 (1997) 998.

\bibitem{22}
D. ter Haar, Elements of Statistical Mechanics, Reinehart, 
New York, 1954.

\bibitem{23}
L.D. Landau, E.M. Lifshitz, Statistical Mechanics, Pergamon,
Oxford, 1958.

\bibitem{24}
J.E. Mayer, M.G. Mayer, Statistical Mechanics, Wiley, New York, 1977.

\bibitem{25}
O. Penrose, Rep. Prog. Phys. 42 (1979) 1937.

\bibitem{26}
V.I. Yukalov, Mosc. Univ. Phys. Bull. 25 (1970) 49.

\bibitem{27}
V.I. Yukalov, Mosc. Univ. Phys. Bull. 26 (1971) 22.

\bibitem{28}
W.H. Zurek, Prog. Theor. Phys. 89 (1993) 281.

\bibitem{29}
B.V. Chirikov, O.V. Zhirov, J. Exp. Theor. Phys. 120 (2001) 214.

\bibitem{30}
G.M. Zaslavsky, Physics Today, N8 (1999) 39.

\end{thebibliography}
\end{document}